\documentclass[journal]{vgtc}                     
\graphicspath{{figures/}{pictures/}{images/}{./}} 

\PassOptionsToPackage{warn}{textcomp}  
\usepackage{textcomp}                  
\usepackage{mathptmx}                  
\usepackage{cite}                      
\usepackage{tabu}                      
\usepackage{booktabs}                  

\usepackage{enumitem}
\setlist{nosep}

\usepackage{dirtytalk}
\usepackage[frozencache,cachedir=.]{minted}

\newcommand{\sys}[1]{\textit{AutoProfiler}}
\newcommand{\inlinep}[1]{\textit{StaticProfiler}}

\MakeRobust{\sys}
\setminted{fontsize=\small,baselinestretch=1}

\usepackage[svgnames, table]{xcolor}

\definecolor{e_orange}{RGB}{248, 229, 208}
\definecolor{e_green}{RGB}{220, 233, 213}
\definecolor{e_blue}{RGB}{166, 196, 229}
\definecolor{e_purple}{RGB}{216, 211, 231}
\definecolor{e_pink}{RGB}{230, 210, 219}

\newcommand{\edit}[1]{{#1}}                  

\usepackage{quoting}
\quotingsetup{vskip=2pt}

\onlineid{1517}

\vgtccategory{Research}
\vgtcpapertype{system}

\title{Dead or Alive: Continuous Data Profiling for Interactive Data Science}

\author{\authororcid{Will Epperson}{0000-0002-2745-4315}, \authororcid{Vaishnavi Gorantla}{0009-0007-1764-6763}, \authororcid{Dominik Moritz}{0000-0002-3110-1053}, \authororcid{Adam Perer}{0000-0002-8369-3847}}
\authorfooter{
  \item All authors are with Carnegie Mellon University. Emails: willepp@cmu.edu, vaishnag@andrew.cmu.edu, domoritz@cmu.edu, adamperer@cmu.edu.
}

\shortauthortitle{Epperson \MakeLowercase{\textit{et al.}}: Dead or Alive: Continuous Data Profiling for Interactive Data Science}

\abstract{
Profiling data by plotting distributions and analyzing summary statistics is a critical step throughout data analysis.
Currently, this process is manual and tedious since analysts must write extra code to examine their data after every transformation. 
This inefficiency may lead to data scientists profiling their data infrequently, rather than after each transformation, making it easy for them to miss important errors or insights. 
We propose \textit{continuous} data profiling as a process that allows analysts to immediately see interactive visual summaries of their data throughout their data analysis to facilitate fast and thorough analysis.
Our system, \sys{}, presents three ways to support continuous data profiling: (1) it automatically displays data distributions and summary statistics to facilitate data comprehension; (2) it is live, so visualizations are always accessible and update automatically as the data updates; (3) it supports follow up analysis and documentation by authoring code for the user in the notebook.
In a user study with 16 participants, we evaluate two versions of our system that integrate different levels of automation: both automatically show data profiles and facilitate code authoring, however, one version updates reactively (``live'') and the other updates only on demand (``dead''). 
We find that both tools, dead or alive, facilitate insight discovery with 91\% of user-generated insights originating from the tools rather than manual profiling code written by users.  
Participants found live updates intuitive and felt it helped them verify their transformations while those with on-demand profiles liked the ability to look at past visualizations.
We also present a longitudinal case study on how \sys{} helped domain scientists find serendipitous insights about their data through automatic, live data profiles.
Our results have implications for the design of future tools that offer automated data analysis support.
}


\teaser{
  \centering
  \includegraphics[width=.97\linewidth]{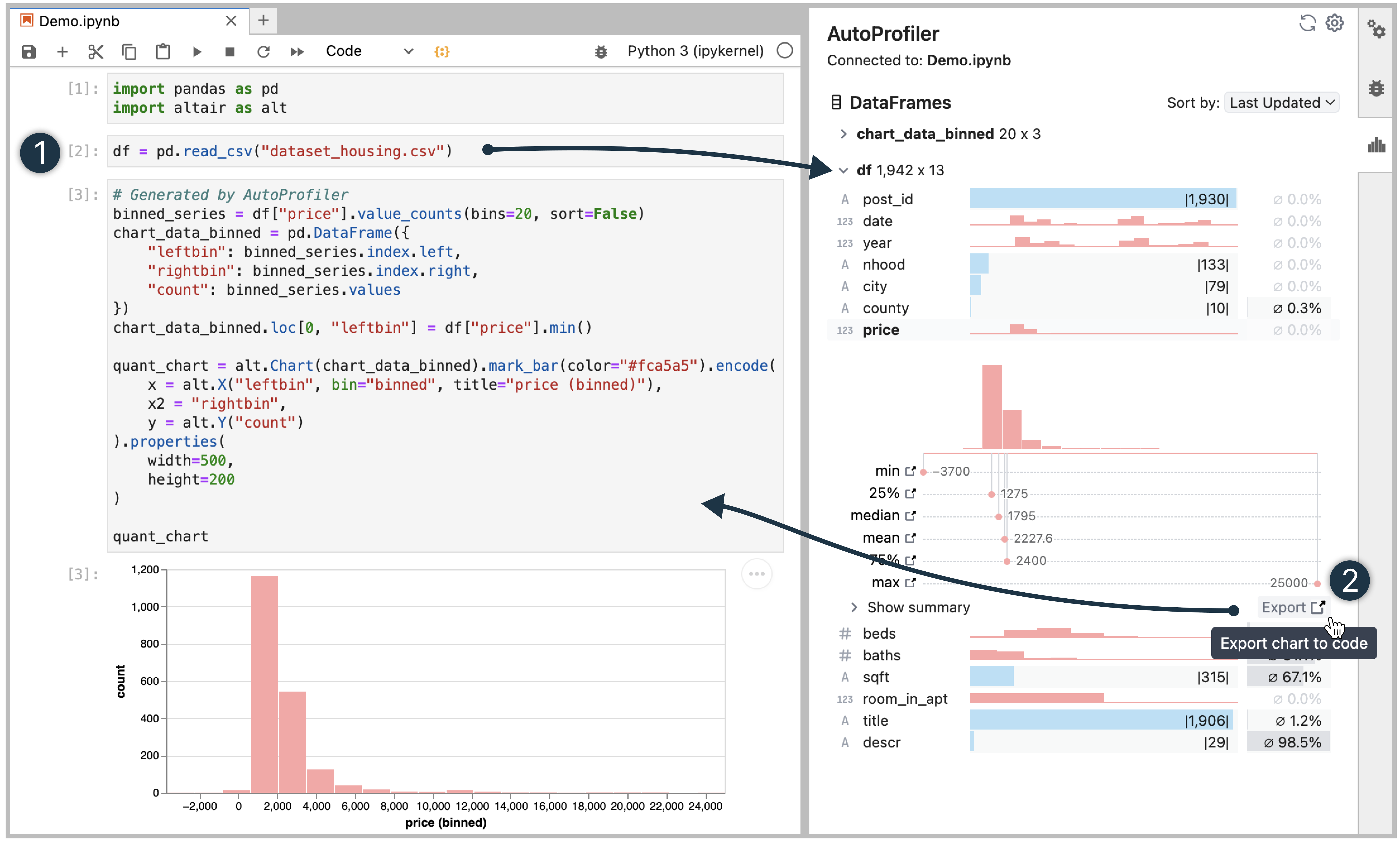}
   \caption{
   \edit{In \sys{}, data profiles update whenever the data in memory updates and are sorted with the last updated dataframes at the top. }
   \edit{In this example, a user has \textbf{(1)} loaded a dataframe about housing prices and sees the profile for \mintinline{Python}{df} in the sidebar.}
   \edit{\textbf{(2)} The user then investigates the price column and exports a chart to code so they can persist this chart and tweak the code for follow-up analysis.}
  }
  \label{fig: teaser}
}

\begin{document}

\firstsection{Introduction}

\maketitle

In recent decades, data analysis is no longer bottlenecked by the technical feasibility of executing queries against large datasets, but by the difficulty in choosing where to look for interesting insights~\cite{Bailis2016PrioritizingAI}.
Interactive programming environments such as Jupyter notebooks help since they support fast, flexible, and iterative feedback when programming with data~\cite{Alspaugh2019FutzingAM, perkel_2018}.
However, while these coding tools were designed to track the state of program execution and variables for debugging, they were not inherently designed to track how data is manipulated and transformed.
This forces users to manually make sense of and write additional code to explore their data.

Exploratory Data Analysis (EDA) is critical to understanding a dataset and its limitations and is a common task at the beginning of a data analysis \cite{Tukey1980WeNB, Wongsuphasawat2019GoalsPA}.
Yet the manual effort required to construct data profiles for EDA takes up a significant part of data analysts' time: recent surveys of data scientists show that they spend almost 50\% of their time just cleaning and visualizing their data \cite{anacondaSODS20202}. 
Since data profiling is so time intensive, it is easy for users to skip over important trends or errors in their data.
This can lead to negative downstream consequences when this data is used for modeling and decision-making \cite{Sambasivan2021EveryoneWT}.
In particular, many data quality issues are potentially silent: models will still train or queries will execute, but the results will be incorrect \cite{Hellerstein2008QuantitativeDC}.
For example, in the data profile of apartment prices in  \autoref{fig: teaser} we can see that some apartment prices have negative values.
If these values are not addressed, analyses or models that use this data may lead to wrong decisions.

We propose \textit{continuous} data profiling as a process that allows analysts to immediately see interactive visual summaries of their data throughout their data analysis to facilitate fast and thorough analysis.
To explore how automated tools can best support continuous data profiling, we have built a computational notebook extension \sys{} that tightly integrates data profiling information into the analysis loop.
\sys{} maintains the advantages of the interactive notebook programming paradigm, while giving users immediate feedback on how their code affects their data.
This tightens the feedback loop between manipulating data and understanding it during data programming.

We explore three main features in \sys{}.
First, it automatically displays profiling information about each dataframe and column to facilitate data understanding.
By showing data distributions and summaries, \sys{} jump-starts a user's EDA.
Second, when the data in memory updates, the profiling information updates accordingly.
``Live'' updates in user interfaces have been shown to reduce iteration time~\cite{Maloney1995DirectnessAL}; with \sys{} we apply this concept to data profiling to understand how it helps facilitate data understanding.
Third, although \sys{} eliminates the repetitive work of authoring data profiling code, users still need to be able to conduct flexible follow-up analysis and persist interesting findings in their notebook~\cite{ruleExplanation2018}.
\sys{} supports this by authoring code for the user through code exports to help users quickly select subsets, find outliers, or author charts.

We present two complimentary evaluations of \sys{}.
In a user study with 16 participants, we evaluate two levels of automated assistance to see how different versions of the tool help users find errors and insights in their data.
Half of the participants used \sys{} (a ``live'' profiler) and the other used a version that presents the same information but in a static, inline version (which we denote as ``dead'').
In this evaluation, we found that users experience similar benefits from both versions of the tool, ``dead'' or ``live'', and generate 91\% of findings from the tools as opposed to their own code.
Participants found live updates intuitive and felt it helped them verify their transformations while those with static profiles liked the ability to look at past visualizations.
Furthermore, participants described how the systems sped up their analysis and exports facilitated a more fluid analysis.
In our second evaluation, we conducted a long-term deployment of \sys{} with domain scientists to use the system during their analysis.
These users described how the ``live'' system enabled them to find and follow up on interesting trends and how \sys{} facilitated serendipitous discoveries in their data by plotting things they might not have checked otherwise.
We discuss how future automated assistants can build on \sys{} to augment data programming environments.
In summary, our paper makes the following contributions:
\begin{enumerate}
    \item We demonstrate the benefits of continuous data profiling with \sys{}, which supports data programming with automatic, live profiles and code exports.
    \item We evaluate this tool in a controlled study and demonstrate how continuous profiling helps analysts discover insights in their data and supports their workflow. 
    \item We also present a longitudinal case study demonstrating how \sys{} leads to insights and discoveries during daily analysis workflows for scientists. 
\end{enumerate}
\section{Related Work}

Our work builds on prior literature on assisted data understanding, live interfaces, and linking GUI and code interfaces.

\subsection{Data understanding is critical yet cumbersome}
Understanding data and its limitations has long been an important, but often overlooked, part of analysis.
Tukey was an early advocate for plotting distributions and summary statistics to get to know your data before confirmatory analysis (hypothesis testing) began~\cite{Tukey1980WeNB}.
Current best practices taught in introductory statistics courses still emphasize the importance of starting analysis with summaries of individual columns, such as distributions and descriptive statistics, before moving on to plot combinations of columns or investigating correlations~\cite{seltman_2018}.
Recent research has highlighted how with the increasing emphasis on developing AI models, people often undervalue data quality leading to negative downstream effects~\cite{Sambasivan2021EveryoneWT}.
Multiple surveys of production data scientists describe the difficulty and time spent on data understanding, profiling, and wrangling \cite{kandelEnterprise2012, anacondaSODS20202, kimDataSciSoftwareTeams2018}.
For example, a recent Anaconda foundation survey described that data scientists self-reported spending almost 50\% of their time on data cleaning and visualization~\cite{anacondaSODS20202}.

Data understanding is difficult because of a variety of factors, including that data updates quickly in production environments, so automated methods and alerts have a high number of false positives \cite{Shankar2022OperationalizingML}, current popular tools require manual data exploration and become messy \cite{perkel_2018}, and as datasets have grown, there are a large number of issues to check for.
Prior systems in the visualization community have addressed parts of this space such as comparing data over time as models are trained on subsequent data versions \cite{Hohman2020UnderstandingAV} or methods for cleaning up notebooks during analysis~\cite{headMessesNB2019}.
However, more work is needed to understand how tools can facilitate discovering data and potential quality issues before they propagate to downstream models or analyses.

\subsection{Prior assisted and integrated EDA tools}
Prior visualization systems aim to automate the visual presentation of data to speed up data understanding.
In general, this automation helps alleviate the burden of specifying charts so that users can focus more on insights rather than how to produce a specific chart \cite{Heer2019AgencyPA}.
Some systems automate visual presentation and then rank charts according to metrics of interest such as high correlation \cite{Demiralp2017ForesightRD}, charts that satisfy a particular pattern in the data ~\cite{Siddiqui2016EffortlessDE}, or contain attributes of interest \cite{2016-voyager}.
Closely related to our work is the Profiler system, which checks data for common quality issues such as missing data or outliers, and presents potentially interesting charts to the user \cite{kandelProfiler2012}.

However, many of these systems exist in standalone tools, making them difficult to integrate into flexible data analysis workflows in programming environments like Jupyter notebooks~\cite{Alspaugh2019FutzingAM}.
Other systems have explored how to integrate visualization recommendations in the notebook programming context as well \edit{through visualization callbacks, libraries, embedded widgets, and similar notebook search~\cite{onoInteractiveDVInNB2021, EDAssistant2023}}.
Lux \cite{Lee2021LuxAV} and other open source tools \cite{bamboolib, pandas-profiling, dataprep, sweetviz, pandasgui} show EDA information on demand for individual Pandas dataframes. 
While Lux uses \say{always on} visualization recommendations to overwrite the default table view for pandas dataframes, users must still ask for visualizations by calling a dataframe explicitly.
\textit{Diff in the Loop} \cite{wangDITL2022} presents a paradigm for automatically visualizing the differences between dataframes after each step in an analysis.
Although these prior systems use automatic visualization, they still require the user to manually ask for this information after each data update and often present an abundance of information that can be difficult to compute in reactive times and for users to parse quickly.
With \sys{}, we explore the benefits and design constraints around coupling automatic visualization with live updates and code authoring on the user's behalf.

\subsection{Liveness in user interfaces}
Fast iteration on data and models is a key element to effective data science~\cite{fisher2012interactions, Shankar2022OperationalizingML}.
The fast, incremental feedback that users receive in Jupyter notebooks is part of the popularity of the platform~\cite{perkel_2018, epperson2022strategies}, yet the default presentation of data feedback in Jupyter is limited to a handful of rows.
``Liveness'' in user interfaces reduces iteration time through reactive updates~\cite{Maloney1995DirectnessAL}, such as in spreadsheets~\cite{hermansSpreadsheetsRCode2016}. 
Prior studies of liveness in data science tools have compared live interfaces to REPL (read-eval-print-loop) interfaces like Jupyter and found users like the responsiveness and clean coding that live interfaces afford~\cite{delineLiveDS2015}.
Inspired by the affordances of live, reactive updates, \sys{} evaluates how automatically updating data profiles after a user changes their data can help reduce iteration time during analysis.
When using \sys{} in Jupyter, users must still explicitly execute their code to manipulate the data, thus it is not a completely ``live'' environment.
However, data profiles reactively update when data changes.

\subsection{Linking code and GUI interactions}
There is a tradeoff between tools that support using code to interact with data or direct manipulation.
Programming languages are flexible and expressive, yet GUIs are responsive and easy to use \cite{Alspaugh2019FutzingAM}.
Prior systems in the notebook setting have bridged this gap by writing interactions with a chart \cite{b2Wu2020} or widget \cite{mageKery2020} back to the notebook automatically.
This allows users to reuse analysis code and preserves the steps of their analysis. 
Selection exports in \sys{} serve a similar purpose of facilitating drill down into rows of interest in a dataset.
Our code authoring approach differs from prior systems since we only write code to the notebook explicitly when the user asks, rather than implicitly after every interaction to avoid polluting the user's notebook.

Beyond their flexibility, programming languages remain popular for data science because they allow users to reuse old analysis code for new purposes \cite{keryVariolite2017}, or use analysis ``templates'' to help users go through the same steps of analysis for similar tasks \cite{epperson2022strategies}.
\sys{}'s template exports serve a similar purpose to author code in the notebook and support follow-up analysis for tasks like customizing a plot, doing outlier analysis, or investigating duplicates.

\section{Design Goals}
We developed the following design principles to inform our system requirements and design:

\begin{enumerate}
    \item [\textbf{G1:}] \textit{Automatic \& Predictable:} Basic data profiling information should be visualized automatically without any need for extra code in a consistent manner. 
    \item [\textbf{G2:}] \textit{Live: } When the data updates, so should all visualizations of it. This prevents ``stale'' data visualizations in a notebook and allows data profiles to be accessible throughout an analysis.
    \item [\textbf{G3:}] \textit{Non-intrusive:} Since users are writing code to interact with their data, automatic visualization should not interfere with their flow. 
    \item [\textbf{G4:}] \textit{Initiate EDA:} Data profiles should present a starting point for understanding each column, which can inform follow-up analysis. 
    \item [\textbf{G5:}] \textit{Persistence:} Tools should support writing findings to the notebook to enable reproducible and shareable analysis.
\end{enumerate}

\textbf{G1} and \textbf{G2} were motivated by the manual EDA which is the current status quo in notebook programming.
We build on prior techniques in live interfaces~\cite{Maloney1995DirectnessAL} and automatic visualization~\cite{Lee2021LuxAV, Heer2019AgencyPA} to speed up the data profiling process and enable continuous data profiling.
This eliminates the need to write repetitive profiling code to understand dataframes after each update.
Importantly, we show the same profiling information for each type of column and visualize the data ``as is'' in order to facilitate finding issues (\textbf{G1}).
With live updates, we situate our profiler alongside the programming environment rather than inline (\textbf{G3}) so that it does not take programmers out of their analysis \textit{flow}~\cite{spaceDevProdForsgren2021}.
This also helps declutter the programming environment since most preliminary visualization can be done in the sidebar.
We make the design choice to show univariate profiling information to help users jump-start their EDA process (\textbf{G4}).
Previous profiling systems often require scrolling to look through multiple pages of charts~\cite{Lee2021LuxAV, pandas-profiling}, making it hard to find interesting problems or insights.
Our goal is to facilitate rapid data understanding with data profiles, then allow users to do further custom analysis by handing off their analysis back to code through exports.
Code exports also facilitate saving findings such as charts or code snippets to the notebook so that notebooks can be shared and reproduced (\textbf{G5}), a core goal in notebook data analysis~\cite{ruleExplanation2018}.

\begin{figure*}[!ht]
    \centering
    \includegraphics[width=\linewidth]{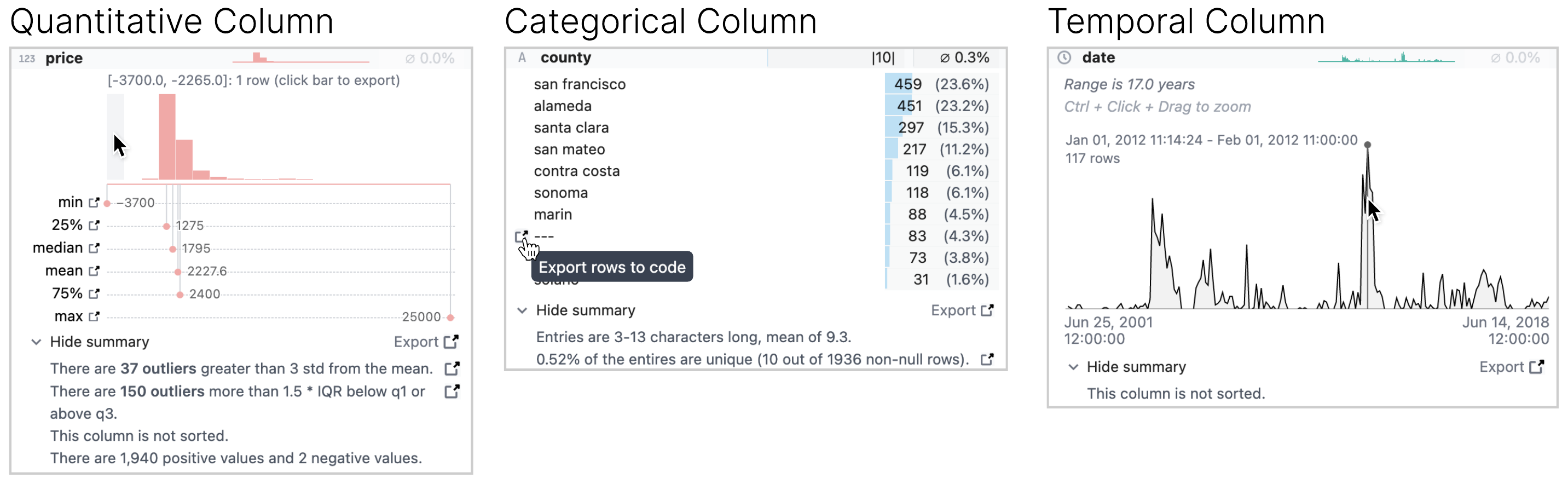}
    \caption{\sys{} shows distributions and summary information depending on the column type. For \textbf{quantitative} columns, we show a binned histogram along with summary statistics. On hover, the user can see the count in each bin or export the selection to code. We also show a summary with extra information like potential outliers that can be exported to code. For \textbf{categorical} columns like strings or boolean values, we show up to the top 10 most frequent values. On click, the selection can also be exported to code. For \textbf{temporal} columns, we show the count of records over time and the range of the column.}
    \label{fig:column_details}
\end{figure*}

\section{Continuous Data Profiling with \sys{}}
\label{sec: AP-system-details}

\sys{} provides data analysts rapid feedback on how their code affects their data to speed up insight generation.
The system fits into a common existing workflow for analysis: using Pandas in Jupyter. 
Pandas is the most popular data manipulation library in Python, with millions of downloads every week~\cite{pandas-dev}.
Likewise, computational notebooks in Jupyter have become the tool of choice for data science in Python~\cite{perkel_2018}.
\sys{} focuses on Pandas users in Jupyter with the goal that features that support this workflow will generalize to other dataframe libraries such as Polars \cite{polars} or Arrow \cite{pyarrow}, as well as other notebook programming environments.
The \sys{} system has three core features that enable continuous data profiling: automatic visualization (\autoref{sec: system:auto-eda}), live updates (\autoref{sec: system:auto-updates}), and code exports (\autoref{sec: system:exports}).

\subsection{AutoProfiler shows EDA automatically}
\label{sec: system:auto-eda}

\sys{} detects all Pandas dataframes in memory and presents them in the sidebar of the notebook.
Each dataframe profile can be shown or hidden, along with more information about each column.
This allows users to drill down into dataframes and columns of interest to see more information, providing details on demand.
By situating \sys{} in the sidebar it also allows users to simultaneously look at both summary data profiles of their data in \sys{} and the default instance view inline from Jupyter.

We use the Pandas datatype of the column to show corresponding charts and summary information.
We categorize the Pandas datatypes into semantic datatypes of numeric, categorical, or timestamp columns similar to previous Pandas visualization systems~\cite{Lee2021LuxAV, solasEpperson22}.
Column profiles for each of these three data types are shown in \autoref{fig:column_details}.
Each column profile has three core components:
\begin{enumerate}
    \item \textit{Column Overview} which contains the name, data type, a small visualization, and the percentage of missing values.
    \item \textit{Column Distribution} which is shown by clicking on the overview to reveal a larger, interactive visualization of column values.
    \item \textit{Column Summary} that has extra facts about a column such as the number of outliers or duplicate values.
\end{enumerate}
The overview, distribution, and summary shown depend on the data type of the column.
Furthermore, the distribution and summary can be toggled on and off to show more details on demand \cite{schneidermanDetails1996}.
This is important for large dataframes with many columns, or when there are many dataframes in memory to prevent unncessary scrolling.
Many visual elements show hints on hover to further prevent visual clutter, providing further details on demand.
Our core charting components were adapted from the open-source Rill Developer platform which shows data profiles for SQL queries~\cite{rillGithub}.
We use the same visualizations in \sys{} with extra summary information and linked interactions to connect the profile to the notebook.

\textbf{Quantitative Columns: } For quantitative columns like integers and floats, we show a binned histogram so that users can get an overview of the distribution of the column.
This histogram is shown in the column overview as a preview; a larger and interactive version is presented upon toggling the column open.
On hover, users can see how many points are in each bin. 
We also show numerical summary information like the min, mean, median, and max of the column.
This is similar to what is presented in the \mintinline{Python}{describe()} function in Pandas to give a numeric summary of a column.
In \autoref{fig:column_details} (left), we demonstrate this information for a price column where we can see that some of the prices in this distribution are negative, a potential error that should be inspected during analysis.

If users want to see more information, they can toggle the summary to see potential outliers, whether the column is sorted, and the number of positive, zero, and negative values.
We use two common heuristics to detect outlier values. 
The first is if a value is greater than 3 standard deviations from the mean; the second is if a point falls outside of $1.5 * IQR$ away from the first or third quartile.
Both forms of outlier detection code can be \textit{exported to code} which allows users to investigate potential outliers more or change these thresholds for classifying the outliers with their code manually.

\textbf{Categorical Columns: } For categorical or boolean columns, we first show the cardinality of the column in the overview to let users understand the total number of unique values.
Once toggled open, the distribution view shows the frequency of the top 10 most common values.
This is similar to the commonly used \mintinline{Python}{value_counts()} function in Pandas which shows the count of all unique values.
In the categorical summary, we show extra information about the character lengths of the strings in the column along with a more detailed description of the column's uniqueness.
This uniqueness fact can be exported to code which lets users inspect duplicated data points.
Once again, users can export a selection to code in the notebook to quickly filter their dataframe. 
For example, in \autoref{fig:column_details} (center) we show the information for the categorical column ``county''.
This column has some default values of \mintinline{Python}{"---"} that seem like an error, so a user can click ``Export rows to code'' to have the code \mintinline{Python}{df[df.county == "---"]} written to their notebook and can investigate these rows further.
Once this new code is written to the notebook, the user can look at this subselection in \sys{} or with their own Pandas code.

\textbf{Temporal Columns: } Our last semantic data type is for temporal columns, where we also show a distribution overview so users can see the count of their records over time.
In the larger distribution view, users can hover over this chart to see the count of values at a particular point in time.
We also show the range of the column and if the column is sorted or not.
Users can drag over a selection of the column to zoom into the time range more in the visualization.
We plan on adding selection exports to temporal columns in the future.
In \autoref{fig:column_details} (right), we show the profiling information for a date column where a user can observe that the records in their dataset span 17 years, however are not evenly distributed with large spikes in certain years such as early 2012.

\subsection{Live Data Profiles}

\begin{figure}[h]
    \centering
    \includegraphics[height=10cm, keepaspectratio]{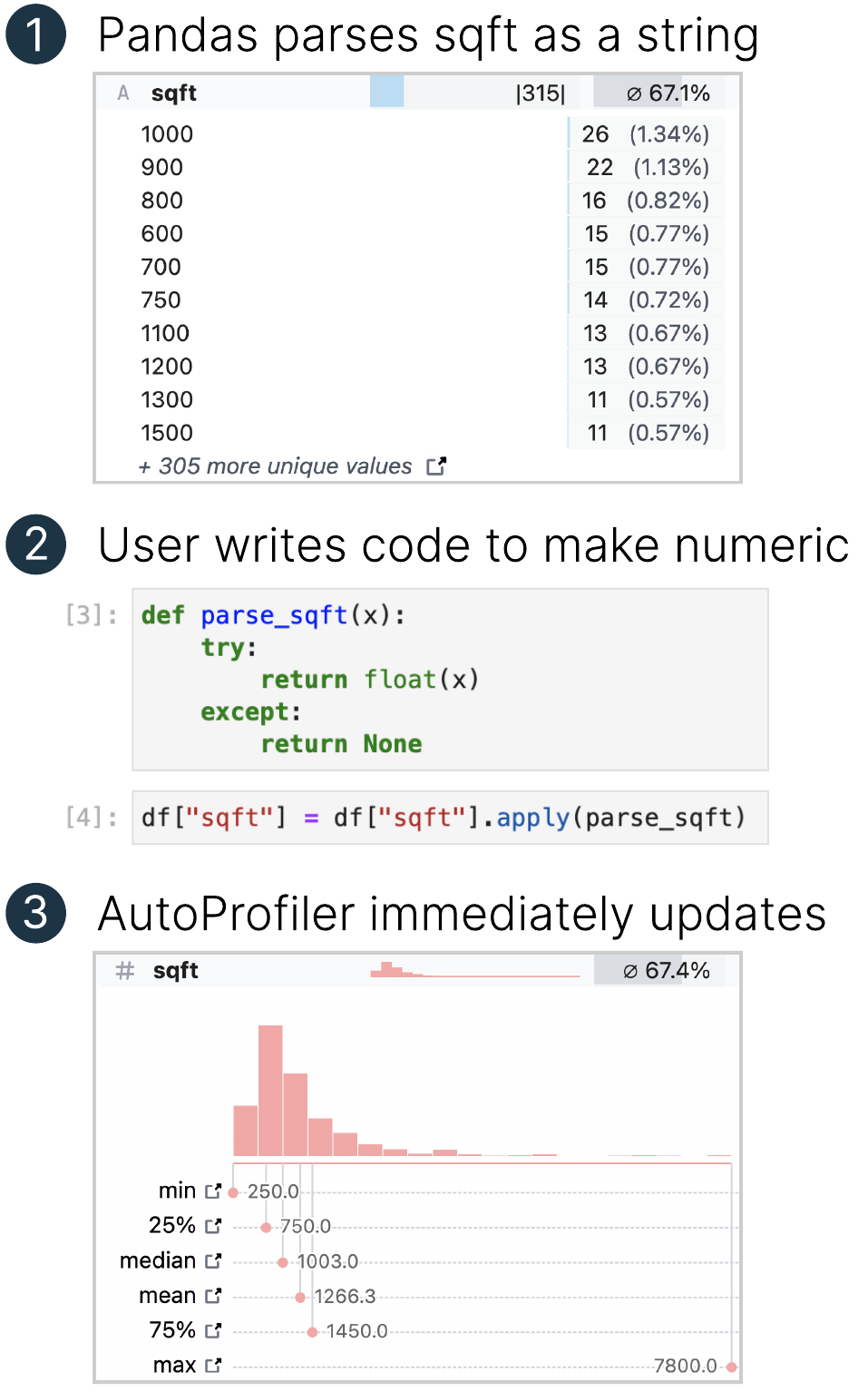}
    \caption{\sys{} updates the data profiles shown as soon as the data updates. 
    In this example, Pandas parses the sqft column as a string type since some of the values initially have strings in them. Once the dataframe \mintinline{Python}{df} updates in memory, \sys{} will update the profile shown. 
    This way the user can see their transformation was successful, inspect the distribution of sqft, and even notice that the number of nulls increased by 0.3\% after this parse. 
    }
    \label{fig:ap_update}
\end{figure}

Beyond showing useful data profiling information just once, \sys{} updates as the data in memory updates.
Once a new cell is executed, \sys{} recomputes the data profiles for all Pandas dataframes in memory and updates the charts and statistics as necessary in the interface.
\edit{With live updates, \sys{} always shows the current state of all dataframes currently in memory in the notebook, allowing users to quickly verify if transformations have expected or unexpected effects on their data.}
\autoref{fig:ap_update} shows this update when a string column is parsed to numeric.
Here, Pandas initially parses this column as an object data type but when the user turns the column into an integer the distribution and summary information is updated.
Live updates help users verify a wide range of transforms.
For example, after updating the types of columns, applying filters, or dropping ``bad'' values.

\sys{} has several UI elements to help users track and assess changes after updates.
The first is that when a user hovers over a column in any dataframe, if other dataframes have columns with the exact same name they are highlighted.
For example, if a user takes the dataframe \mintinline{Python}{df}, filters it to \mintinline{Python}{df_filtered}, and then hovers on the Price column the linked highlights help the user make a visual connection between the two Price columns.
With automatic dataframe detection and visualization, there can potentially be many dataframes in memory as users manipulate their data over an analysis.
\sys{} supports sorting dataframe profiles to find those of interest.
By default, the most recently updated profiles are shown at the top of the sidebar.
A user can also sort alphabetically by the dataframe name.
Furthermore, users can pin any profile so that it always appears at the top of the sort order.

Dataframe profiles are typically only shown for dataframes explicitly assigned to a variable with one exception: if the output from the most recently executed cell is a Pandas dataframe we will compute a profile for it with the name ``Output from cell [5]''.
On the next cell execution, these temporary profiles are removed.
This fits into a common notebook programming workflow where users display their dataframe after making a transformation to see how the data has changed.

\label{sec: system:auto-updates}

\subsection{Exports to code}
\label{sec: system:exports}
In addition to interactive data profiles, \sys{} assists users in authoring code. 
\sys{} facilitates code creation in two ways: \edit{\textit{selection} and \textit{template} code exports}. 
\edit{For both of these, a user clicks on a button or part of a chart and \sys{} writes code for them in the notebook below the user's currently selected cell.
All code export snippets are pre-built into \sys{} and produce the same code snippet for each task with the dataframe and column names filled in so the code is ready to execute in the notebook.}

\edit{Selection and template exports only differ in the kind of code they produce.}
\textit{Selection exports} allow users to export selections from charts to help them filter their data, as mentioned in \autoref{sec: system:auto-eda}.
For example, \autoref{fig:column_details} (left and center) demonstrates how a user can export selections from categorical and numeric charts to quickly filter their data.
This helps users more quickly iterate on ideas during analysis to spend less time writing simple code and proved very popular in our user study.

\sys{} authors more complex code like charts or code to detect outliers with \textit{template exports}. 
Code exports for these tasks are still relatively simple, only exporting up to 10 lines of code.
However, this saves users from having to remember how to author a chart themselves or compute outliers.
Users can then easily edit this code, for example to customize their visualization or change the threshold for an outlier.
Prior work has discovered how data scientists often re-use snippets of code across analyses to help them speed up their workflows~\cite{epperson2022strategies, keryVariolite2017}.
\sys{}'s exports serve as a form of these pre-baked ``templates'' for analysis steps.
The other benefit of this type of export is that it helps preserve analysis in the notebook in the form of code, which supports more replicable analyses in notebooks, a common goal~\cite{Pimentel2019ALS}.

This linking between analysis in a visual analytics tool and notebook code has been introduced in previous systems such as Mage \cite{mageKery2020} and B2 \cite{b2Wu2020}.
Our goal here is similar: to support tight integration between GUI and code.
However, our approach differs slightly in that we only write code to the notebook when the user \textit{explicitly} clicks a button to prevent polluting the user's working environment.

\subsection{Implementation and Architecture}

\begin{figure}[h]
    \centering
    \includegraphics[width=\linewidth, keepaspectratio]{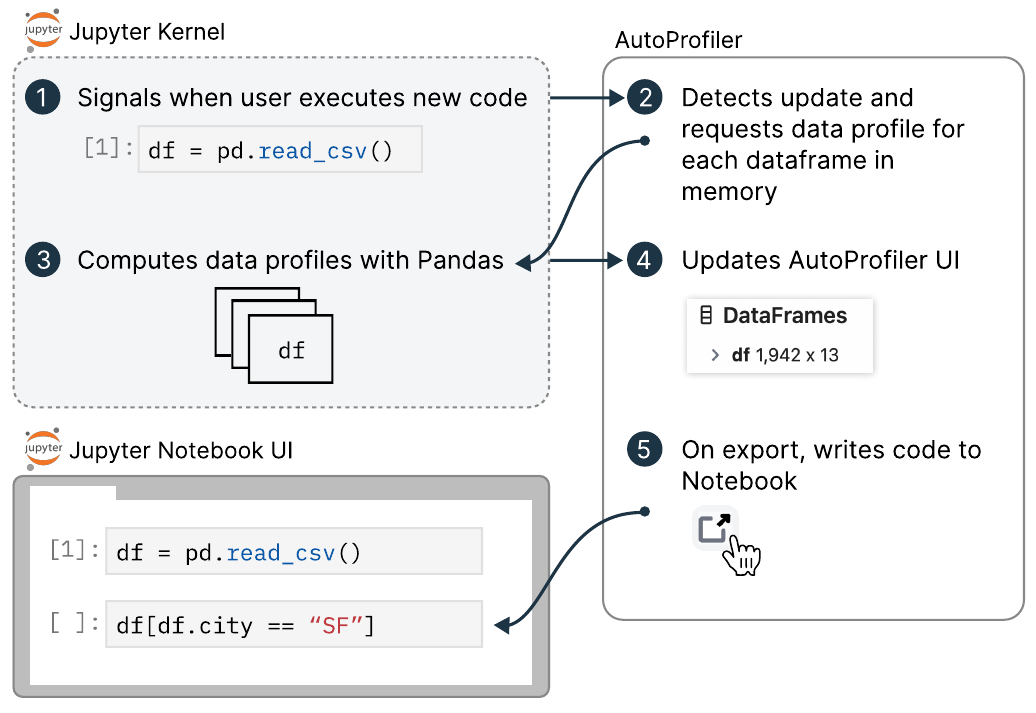}
    \caption{\sys{} \edit{profiling workflow}. Data profiles are computed reactively when a user executes new code. Profiling is done in the kernel to speed up performance and avoid serializing the entire dataframe.
    }
    \label{fig: ap_arch}
\end{figure}

\sys{} is built as a Jupyter Lab extension to augment a normal interactive programming environment with a data profiling sidebar.
\autoref{fig: ap_arch} \edit{shows the components involved in a example live update loop}.
When a user executes new code, the kernel sends a signal that a cell was executed (step 1).
\sys{} then interacts with the kernel to get all variables that are Pandas dataframes, and requests data profiles for each of these variables (steps 2 - 4).
When a user requests to export code, a new cell is created with the code (step 5).
This is only a UI interaction, and when the user executes the generated cell, the update loop will trigger again.
\edit{Whenever the kernel is restarted, the dataframes in memory are cleared so the profiles in \sys{} reset.}

As a Jupyter extension, \sys{} can be easily installed as a Python package and included in a user's Jupyter Lab environment.
This easy installation has proven very popular with users of our system.
The frontend code for \sys{} uses Svelte~\cite{svelte} for all UI components.
Our code is open-sourced and available for use\footnote{ \url{https://github.com/cmudig/AutoProfiler}}.

All profiling functions are written in Python and execute code in Pandas.
Pre-binning distributions in python makes serialization faster \edit{to avoid serializing entire dataframes.}
Since our profiling happens in Pandas, the performance of \sys{} generally scales with the capabilities of Pandas.
\edit{Anecdotally, we have used \sys{} during analyses with dataframes with hundreds of thousands of datapoints and updates remain responsive.}

\edit{The scalability of our approach is primarily impacted by two main considerations: the number of columns in each dataframe and number of dataframes in memory.
Pandas can still execute a single query relatively quickly for dataframes with up to millions of datapoints, and we consider a full benchmarking of pandas queries outside the scope of this work.
Since requests to the Jupyter python kernel are currently executed serially, larger requests for dataframes with many columns or more dataframes in memory make updates slower. 
The \sys{} UI is not affected by the size of the underlying data since the queries return binned data counts or summary statistics so the UI remains responsive, it simply takes longer to fetch new data for larger or more dataframes.
We have included several performance tweaks to make \sys{} usable for real workflows.
For example, we do not calculate updates when the \sys{} tab is closed to avoid unnecessary computation.}

\edit{
The scalability of \sys{} can be improved with further engineering.
For example, the requests for profiling queries could be executed in parallel by augmenting the Jupyter kernel.
Furthermore, faster query execution system like DuckDB~\cite{duckdbEfficientPandas2021} can speed up the response on individual queries over pandas.
For particularly large datasets, the distributions and statistics could be estimated from samples.
}
\section{Evaluation: User Study}
\label{sec: user study}

\begin{table*}[h]
    \centering
    \begin{tabular}{lllllrr}
    \toprule
    No. & Type & Category & Origin & Description &  Found & Found with tool \\
    \midrule
    1 & Missing  & Error & Inherent & Small number of missing values in county, beds, title & 56\% & 100\% \\
    2 & Missing  & Error & Inherent  & Mostly missing in baths, sqft, description & 56\% & 100\%\\
    3 & Inconsistent  & Error & Added & City has values that are lower and upper case & 69\% & 91\% \\
    4 & Inconsistent  & Error & Added  & Negative prices & 69\% & 100\% \\
    5 & Incorrect  & Error & Inherent  & Date could be parsed to DateTime format & 63\% & 90\%  \\
    6 & Incorrect  & Error & Added  & County has default values of \mintinline{Python}{"---"} & 81\% & 85\%  \\
    7 & Incorrect  & Error & Added  & Sqft has string values and should be converted to an int & 69\% & 82\%  \\
    8 & Outliers & Error & Inherent & Outliers in sqft & 6\% & 100\%\\
    9 & Outliers  & Error & Inherent & Outliers in price & 44\% & 100\%\\
    10 & Schema   & Error & Added & Duplicate datapoints (duplicate post\_ids) & 38\% & 100\%  \\
    11 & Distribution  & Insight & Inherent  & Room\_in\_apt is almost all 0 & 56\% & 100\% \\
    12 & Scope  & Insight & Inherent & Dataset is only apartments in California & 31\% & 100\%\\
    13 & Correlation  & Insight & Inherent & Inspect correlations between any variable and price & 13\% & 0\%\\
    \midrule
    14 & Distribution & Insight & Inherent & Data is not evenly distributed across years & 38\% & 100\% \\
    15 & Inconsistent & Error & Inherent & Year and date column correspond (ensure consistency) & 19\% & 67\% \\
    16 & Inconsistent & Error & Inherent & The price is not properly extracted from title for some rows & 6\% & 0\% \\
    \bottomrule
    \end{tabular}
    \caption{Description of each of the errors and insights on our ``rubric'' of participant performance.
    We include the percentage of participants that discovered each error/insight, noting that some discoveries were found far more often than others. 
    \edit{As the same information was present in both \sys{} and \inlinep{}, the discovery rate in each condition is largely comparable.}
    The first 13 insights and errors were things we expected participants to discover ahead of time, and the last 3 were valid extra findings discovered by participants. 
    }
    \label{tab: errors in study}
\end{table*}

We demonstrate the effectiveness of \sys{} in two ways.
In this section, we discuss the results of a user study comparing two levels of automation support with \sys{} and in \autoref{sec: case study} we discuss the results of a longitudinal case study of users with \sys{}.

\subsection{Participants}
To evaluate how \sys{} helps data analysts in a sample data analysis task, we recruited Pandas and Jupyter users for a between-subjects user study. 
We recruited 16 participants from social media and our networks who were experienced data analysts. 
Our inclusion criteria required that participants be regular Pandas and Python users.
Our participants had 2 to 12 years of experience doing data science (mean 4.8 years), and were all regular Python and Pandas users who frequently used Juptyer.
The typical participant reported doing data analysis weekly and using Pandas daily, with all participants using Pandas at least monthly.
Our participants worked in a variety of industries including autonomous vehicles, data journalism, and finance with job titles including data analyst, data engineer, post-doc, and researcher.

\subsection{Research Questions}
We had three primary research questions in our user study:
\begin{enumerate}
    \item [\textbf{Q1.}] \textit{Live updates:} Does a profiler with live updates lead to more insights found than one with manual updates?
    \item [\textbf{Q2.}] \textit{Starting point for EDA:} Does automatically providing visual data profiles lead users to write less code, and is this information helpful? 
    \item [\textbf{Q3.}] \textit{Linked code and GUI:} How does code exporting facilitate handoff for follow-up analysis?
\end{enumerate}

These research questions correspond to the main features of our tool. 
We test how different levels of automation support continuous data profiling for \textbf{Q1} by comparing the number of insights found through a profiler with live updates to one that required manual invocation.
With \textbf{Q2}, we explore our design choice of showing a starting point for data profiling.
To answer this question we measure how many insights participants found through our tools versus their own code and their qualitative perceptions of each tool version.
Finally, to answer \textbf{Q3} we measured how often exports to code are used during analysis and participants' perceptions of this feature.

In order to answer these research questions, we ran a between-subjects user study with two versions of our tool. 
We elected for a between-subjects design since data analysis requires time to do well and we found during pilots that having participants analyze two separate datasets was infeasible and the quality of analysis on the second task was significantly worse.
We also noticed a large learning effect in pilot studies when participants analyzed two datasets back to back.

\subsection{\inlinep{}}
In our study, one condition used \sys{} with live profiles, automatic updates, and code exports.
For our other condition, participants used a \textit{static} version of the tool which we call \inlinep{} which requires manual invocation.
\inlinep{} allows us to test how different levels of automation support continuous data profiling.
The interface shows the exact same information as \sys{}, however, it must be called manually with \mintinline{Python}{plot(df)} and does not update automatically with data updates.
The same profiles for each column are presented in an inline interactive widget with the ability to hand off to code in the notebook.
This sort of manual invocation is similar to other Pandas visualization tools in notebooks~\cite{Lee2021LuxAV, bamboolib, pandas-profiling}.
\edit{A screenshot of the \inlinep{} tool is included in the appendix.}

\edit{We compare \sys{} with \inlinep{} rather than other open source tools since \inlinep{} includes largely the same information as other tools but the UI design is the same as \sys{}.}
Our goal with this comparison was to isolate the effects that live updates have on continuous data profiling (\textbf{Q1}) and evaluate \textbf{Q2} and \textbf{Q3} through logs and interviews across both system versions.
\edit{We compare \sys{} to a non-live updating tool, \inlinep{}, instead of a baseline of no tool since participants could write any extra code in the study notebook and did not have to use the tools. 
This allowed us to evaluate how different designs impacted tool use and how a tool augmented a typical programming workflow.  }

\subsection{Procedure and Task}
In both conditions, participants were first shown a demo of the tool version they would be using (\sys{} or \inlinep{}).
Each participant then analyzed the same dataset during the task.
The dataset was a sample of a larger dataset of apartment listings from craigslist~\cite{bayAreaRentsData} with extra ``errors'' added\footnote{Task dataset: \url{https://github.com/cmudig/AP-Lab-Study-Public}}.
The task dataset had 1,942 rows and 13 columns.
\edit{We sampled the dataset to a smaller size so we could be more confident that our rubric covered the majority of important insights and errors in the data.}

We had 13 pre-known insights/errors that we measured to see how well participants could explore the data and find these insights as an inital ``rubric'' of task performance.
Additionally, we included three extra insights and errors that participants found during their exploration.
A detailed description of each insight/error that we measured is in~\autoref{tab: errors in study}.
The categories of errors in this dataset were inspired by prior studies that group dataset errors into common types\cite{kandelProfiler2012}.
Our first 10 dataset errors are issues of missing data, inconsistent data, incorrect data, outliers, and schema violations.
Inconsistent data refers to data with inconsistencies like variations in spelling or units; incorrect data is parsed as the wrong data type or has default values like dashes or empty strings.
In addition to errors that might jeopardize an analysis if not discovered, we also measured how well participants discovered several broader insights in the dataset.
\edit{Building off past definitions of dataset insights as unexpected, qualitative findings rooted in the data \cite{North2006TowardMV}, we broadly considered insights as findings about the data that did not fit into one of the aforementioned error buckets and are important to know before the dataset is used for a downstream task.}
We initially included three general insights such as the scope of the dataset, realizing skewed distributions, and investigating correlations.
While these errors/insights are by no means exhaustive of everything of interest in our dataset, they provide a common ``rubric'' that we could evaluate participants against.
\edit{We consider this rubric indicative of things that \textit{should} be found in a proper EDA of the dataset, regardless of the tool being used.
With the exception of insight 13 about correlations, all of these findings can be seen in the \sys{} or \inlinep{} interfaces.}

Participants were asked to explore and clean the data under the guidance that this dataset was recently acquired by a colleague who wants to build a predictive model of apartment prices.
Participants were asked to clean and produce a report about the dataset in the notebook that would be handed off to their colleague.
Participants were told there were at least 10 errors in this dataset that they should try to find and fix to encourage critical engagement with the data.
They were not told what kind of errors these were or what constituted an error.

Participants were given 30 minutes to explore the data with the tool and asked to think aloud about what they were investigating. 
Participants were asked to write down any insights and findings in their notebooks and voice them aloud.
During their analysis, they were free to look up external documentation and use any other python libraries they thought might be helpful.
Our research team was present if participants had questions about the task overall, however, did not answer questions about the data.
We automatically logged interactions with the tools during the study.
Afterward, we conducted semi-structured interviews with each participant and asked them about how they went about the task and how the tool supported their analysis.
\edit{We examined the findings that participants wrote down in the notebook or voiced aloud from study recordings to quantify how many of the insights on our rubric they had found.}
In Sections~\autoref{sec: results-live}, \autoref{sec: results-speedup}, and \autoref{sec: results-exports} we discuss findings based on these logs and interview data. 

\begin{figure}[h]
    \centering
    \includegraphics[width=\linewidth, keepaspectratio]{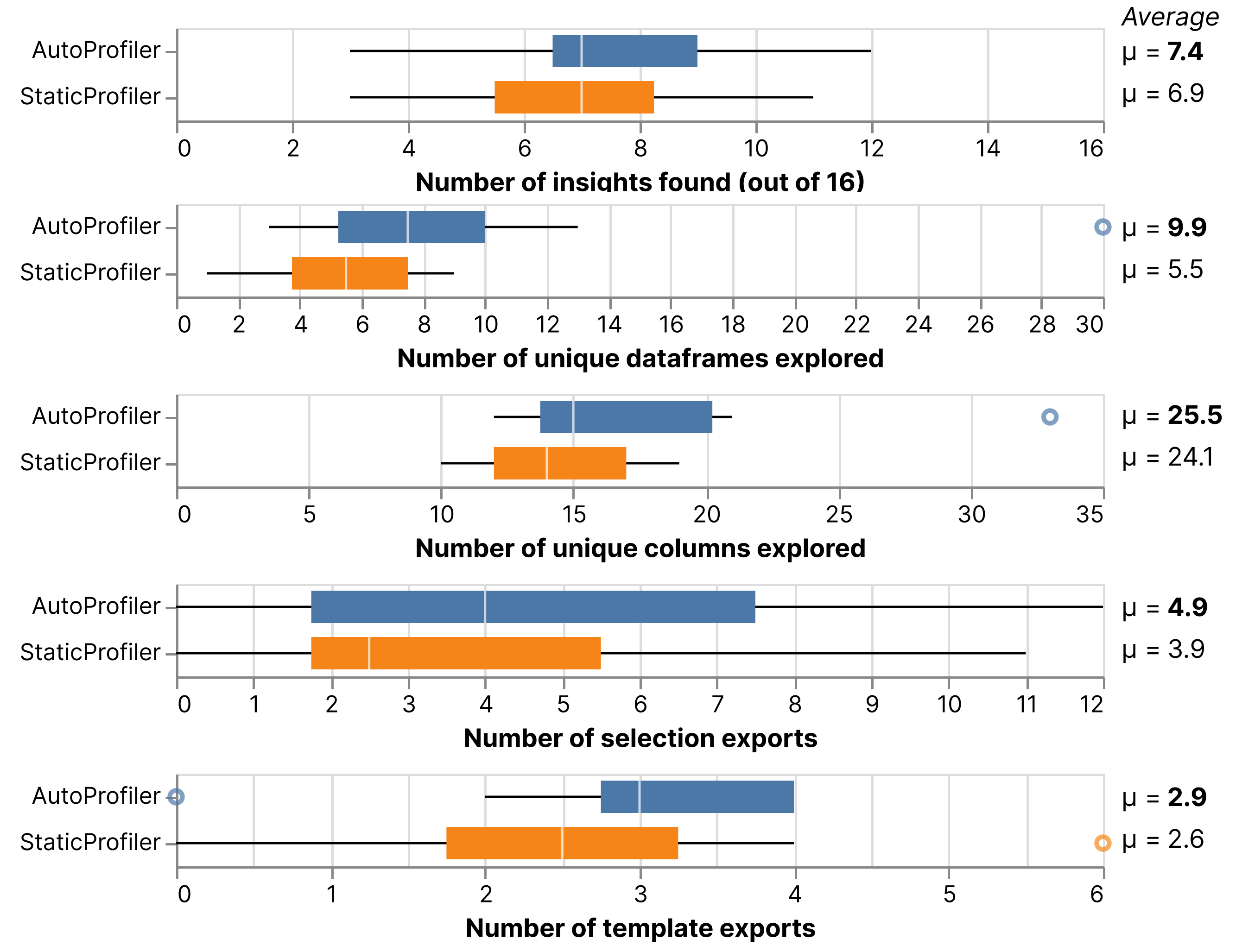}
    \caption{Usage and task performance metrics of \sys{} and \inlinep{} from our user study.}
    \label{fig: metrics_dist}
\end{figure}

\subsection{Live profiles do not lead to more insights but make verification easier}
\label{sec: results-live}

In both conditions, participants found a similar number of insights: on average, 6.9 with \inlinep{} and 7.4 with \sys{} out of the 16 we measured (\textit{P}=0.71).
Therefore, we did not observe more insights found with \sys{} (\textbf{Q1}).
Participants heavily used both versions of the tool as demonstrated by the similar number of unique dataframes and columns explored in \autoref{fig: metrics_dist}.
\edit{We suspected the live updates in \sys{} to encourage \textit{more} tool use which would lead to more insights found but participants found both versions to be helpful during their analysis task, reinforcing the value of automatic visualization.}
Furthermore, live updates may not have made as much of a difference in a controlled lab setup versus a less well-defined analysis outside of the lab setting which we explore in \autoref{sec: case study}.

Participants used both versions of the tools to verify that their code had the expected effect on a dataframe.
For example, we observed participants finding an error through the tool, writing code to fix it, and then checking that their code had the expected effect through the tool. 
We particularly noticed this pattern with users of \sys{}.
For example, P3 noticed error \#3 that the city column contained some cities that were spelled with different casings (``Oakland'' and ``oakland'') with the column detail view.
They then fixed this error by making all the values upper case with their own Pandas code and verified that the top values were all upper case in \sys{}.
As P3 described: 
\begin{quoting}
\say{It was nice to see when I do the upper [casing] and I can just see, oh that worked. When I do the drop duplicates, I can just look and see like, oh that worked. I like that.}
\end{quoting}

We observed this (1) find a dataset error, (2) fix, and (3) verify in the tool loop for many of our participants.
Live updates help facilitate this verification since the updates happen automatically, whereas with the static version of the tool, users would often verify transformations with their own code manually.
As P7 (\inlinep{}) mentioned: \say{I only want [\inlinep{}] when I'm ready for it. Because it does take up some screen space. Like I don't want it like suddenly bumping a bunch of things out of the way.}
\edit{Since \inlinep{} puts visualizations inline in the notebook, multiple invocations can lead to cluttered notebooks.}

Both \sys{} and \inlinep{} also helped participants quickly discover when they had done a transformation \textit{incorrectly}.
For example, P5 used \sys{} to export the outliers for the beds column to code.
However, when they re-assigned their dataframe variable, they assigned \mintinline{Python}{df} to only contain outliers by accident.
With \sys{} they quickly noticed that their dataframe now only contained 12 data points with extreme distributions and were able to fix their error.
We observed this pattern of the tool helping find user errors during four different studies, three of which were using \sys{}.

Using static, inline data profiles is not without its advantages.
For one, several users liked the ability to keep a history of past dataframes in their notebook when they called \mintinline{Python}{plot()} with \inlinep{}.
Although some participants felt this led to potentially cluttered notebooks, it can be useful to scroll back to an earlier version of the data.
This is not possible in \sys{} since the visualizations always show the current dataframe in memory.

\subsection{Automatic visualizations speed up insight discovery}
\label{sec: results-speedup}

Participants found the tools to be useful both as a first step in analysis, but also to help them understand their data after updates and transforms.
We logged interactions during the study and present metrics of interest in \autoref{fig: metrics_dist}.
We measured the unique dataframes explored as the number of unique dataframes toggled open (\sys{}) or called with plot (\inlinep{}).
This metric captures how often a user returns to a dataframe after it updates or explores a new dataframe.
For example, if a user explores \mintinline{Python}{df}, updates it, then explores \mintinline{Python}{df} again we would count this as two unique interactions.
We observed that participants with \sys{} interacted with slightly more dataframes (9.9 vs 5.5), however, this difference was not statistically significant (\textit{P}=0.21).
Over the course of their analysis, participants were on average inspecting data profiles in \sys{} for almost 10 different slices or updates to dataframes.
One of our participants with \sys{} actually interacted with 30 unique dataframes during their analysis.

We also measured the number of unique columns (including updates) that participants interacted with and find that they explore largely the same number of columns in each condition, investigating 25.5 unique columns on average with \sys{} and 24.1 with \inlinep{}. 
Since the original dataset had 13 columns, this indicates that participants were not only interacting with the original data but were returning to the profiles as they updated or filtered their data.
This continuous interaction is the main goal of continuous data profiling.

Overwhelmingly, participants found their insights \textbf{with} the assistance of either tool \edit{rather than by manually writing code to get the same information.}
This means that when a participant said the insight aloud or wrote it down in their notebook, this information was discovered through the tool.
\edit{Across both conditions, an average of 91\% of insights found came from the tool, with a non-significant difference in rates between the two conditions (\textit{P}=1.0).}
This means that on average only 9\% of insights were found by users writing manual pandas code during the study.
This supports that the information contained in the profiles is useful and replicates what participants would have wanted to see anyway without requiring extra code to be written (\textbf{Q2}).
As P14 (\inlinep{}) said \say{it does a lot of the things that I already do, but just in one succinct and easy-to-understand way}.
By presenting this information automatically, the tools saved participants time and prevented them from having to exit their analysis flow to look up external documentation.
As P10 (\sys{}) described: 
\begin{quoting}
\say{I might have known to look for it, but it would have taken me a lot longer to remember how to do it in Pandas.}
\end{quoting}

When data profiling information is more easily accessible it speeds up the entire analysis loop, making it easier to discover more insights in a shorter amount of time while still being thorough. As P9 (\sys{}) described:
\begin{quoting}
    ``I would probably try to do similar things that AutoProfiler suggests [on my own], but it would take a much longer time. Like the amount I did in 30 minutes, if I had to do it without AutoProfiler, would have taken hours. And then since it takes longer, my motivation would go down and my focus would go down. So I feel like I would have found far fewer errors than I could with AutoProfiler.''
\end{quoting}

We found that not all insights were discovered with the same frequency, with discovery rates between 6\% and 81\%.
In \autoref{tab: errors in study} we see that some errors like \#6 were found by 81\% of participants; others like \#8 or \#10 were found by 6\% and 38\%, respectively.
Error \#8 was particularly difficult since the sqft column had to be parsed from a string to an integer (error \#7) to get information about the outliers in the profiles.  
Many participants did not successfully fix this issue during the study time, explaining the low discovery rate.
However, duplicate primary keys (error \#10) was readily discoverable in the interface by looking at the number of unique values in the post\_id column yet few participants found it.
We discuss this usage trend in more depth in \autoref{sec: discussion} about how tools can facilitate users finding information they would have already wanted to investigate, however if they do not know to check for an issue then this information is easily skipped over.

\subsection{Exports facilitate follow-up analysis and learning}
\label{sec: results-exports}

We also measured the number of times that participants exported to code during their analysis.
Every participant used code exports at least once, with the total number of exports ranging from 1 to 16, with a mean of 7.1 exports.
In \autoref{fig: metrics_dist}, we detail the average number of exports between the two tools.
We see similar trends across both conditions, where participants export more selection exports than template exports.
Selection exports refer to exporting a filter from a chart or summary statistic like exporting the selection for \mintinline{Python}{df[df.city == "San Jose"]}.
Although these exports are small, they can help make follow-up analysis easier if a user wants to filter since \say{that's probably the most annoying lines to constantly type is [to] just filter} (P5, using \sys{}).

Template exports refer to code for authoring a chart or getting outliers.
Participants also found this helpful because it helped facilitate tweaking code for follow-up analysis.
When describing their reason for using chart exports, P14 (\inlinep{}) mentioned \say{It's really nice to just quickly be able to like to copy that and use it, and then I could just make some edits to it.}
This answers \textbf{Q3} that exports facilitate faster feedback loops.

Another unexpected benefit of code exports is the ability to actually learn Pandas better and understand what is going on under the hood of the system when it reports a statistic.
As P12 (\sys{}) said succinctly: \say{I’m learning as I’m exploring and it’s saving me time.}
Expanding more, P2 (\inlinep{}) mentioned:
\begin{quoting}
    \say{For the educational perspective, that’s something I didn't expect...specifically, I [exported] the standard deviation and I could see points inside or outside of 3 [std]. When I saw that code I learned that's the way to do that.}
\end{quoting}
The ability to teach users how to do common analysis steps is an exciting aspect of systems that support easily linking code and direct manipulation interactions.

\subsection{Limitations}
\label{sec: limitations}

Our user study is subject to several limitations.
First, subjects were explicitly told to explore and clean their dataset and were given 30 minutes to engage with a brand-new dataset. 
This is a relatively short time span to learn and use a new tool on new data.
We also suspect that the explicit instructions to find errors and write down findings in a report might have encouraged better continuous data profiling practices than what actually happens in real-world settings.
However, these explicit instructions helped us determine which features specifically aid in continuous data profiling and what kind of errors users commonly find or miss.
Another limitation is that participants analyzed a relatively small dataset.
The errors and insights in our dataset were representatives of those found in larger datasets and we believe our findings translate well to other tabular dataset tasks.
Finally, we compared two versions of our tool with different levels of automation to understand how they supported continuous data profiling rather than comparing to a baseline with no tool and view this as an area for future work.
\begin{figure}[h]
    \centering
    \includegraphics[height=9cm, keepaspectratio]{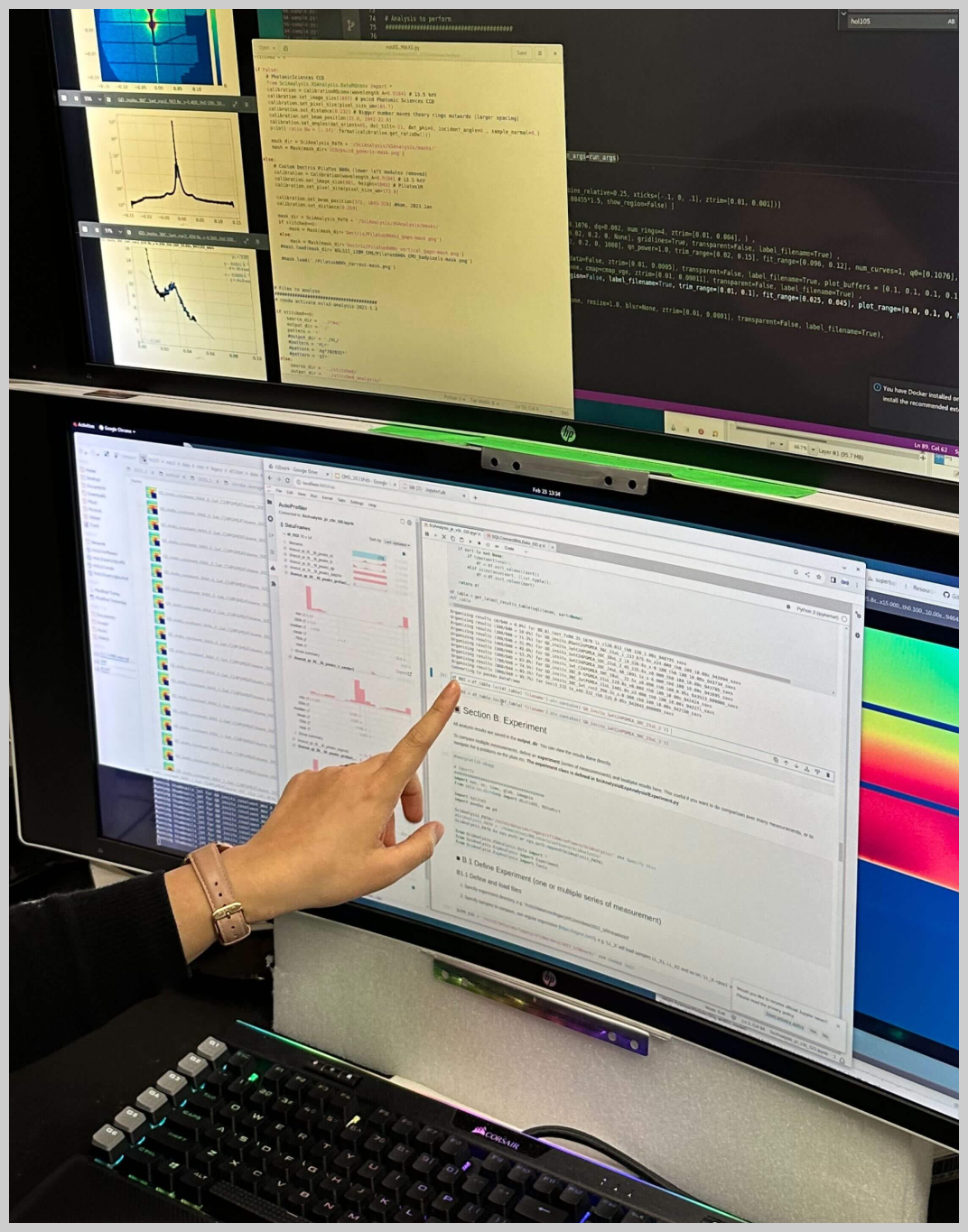}
    \caption{\sys{} integrated into a domain scientist's analysis workflow during our case study. \sys{} is shown on the bottom screen in the Jupyter notebook.}
    \label{fig: case study photo}
    \vspace{-5pt}
\end{figure}

\section{Evaluation: Longitudinal Case Study}
\label{sec: case study}

To address some of the limitations of our user study, we also evaluated how \sys{} helps data scientists in a real world environment by working with domain scientists at a US National Lab to integrate \sys{} into their workflows.
These scientists work with large-scale image data collected from beamline X-ray scattering experiments to understand the properties of physical materials  \cite{Kiapour2014MaterialsDF}.
\edit{
Two different scientists installed \sys{} into their Jupyter Lab environments and used it over a three month period during their analyses as much as they liked.
We were unable to collect log data during this deployment for privacy reasons.
We periodically spoke with the scientists during the deployment to make sure the tool was working.
At the end of the 3-month period, we conducted in-person observations and interviews with the participants where they showed us the notebooks and datasets where they were using \sys{} and we asked about how they used the system, and which features they felt supported their workflows.
}

\edit{As a Jupyter Lab extension, \sys{} fits into the existing workflows of these scientists since they typically did data analysis with Python and had existing libraries for visualizing and manipulating their data.}
\sys{} helped improve two different workflows they have for data analysis.
The first is for monitoring data outputs and quality \textit{while} an experiment is running.
Their experiments last for multiple hours or even days while they collect image readings from a sensor and then process these images into tabular datasets with Python image processing pipelines.
As the scientists describe, during these experiments \say{real-time feedback is important as it shows us whether the experiment is working}.
The participants mentioned how \sys{} improved this type of monitoring since it works with any Python-based analysis and \say{allows [them] to easily notice any anomaly and observe a trend or correlation during experiments.}

The second way the participants used \sys{} was to analyze their results after an experiment completed.
In this scenario, the scientists \say{iteratively sub-selected a relevant set of data, using AutoProfiler as a guide, and then analyzed this subset of data using existing analysis/plotting tools. Thus, AutoProfiler has shown its value in improving data triage, data organization, and serendipitous discovery of trends in datasets}.
In the remainder of this section, we discuss two high-level patterns of use that emerged from interviews with the participants in our long-term deployment. 

\subsection{Finding and following up on trends}
When using \sys{} to analyze their experimental results, our participants expressed how the tool facilitated finding interesting aspects in their data and then diving deeper into those subsets.
In this way, \sys{} facilitated a faster find-and-verify loop during analysis.
The automatic plotting in \sys{} presented interesting plots in their dataset that helped them find subsets to export and explore further such as by running other analysis code to plot the images corresponding to each data point.
They were especially excited about the possibility of incorporating bivariate charts into \sys{} so they would have to use even less of their own analysis code.

\subsection{\sys{} facilitates serendipitous discovery}
The scientists used the live version of \sys{} that updates whenever their data changes.
They mentioned that the combination of all three features (automatic visualization, live updates, and code authoring) supported one another to lower the friction of their data analysis and were not enthusiastic about using versions of the tool without all of these features (such as in \inlinep{}).
Furthermore, the participants mentioned that using \sys{} helped them discover trends or errors they might not have noticed otherwise:
\begin{quoting}
``One of the things that I very often notice is if the histogram is completely flat. That means that either all the numbers are exactly the same, or that it's some sort of sequential number. Sometimes that's what I'm expecting, so great. But sometimes, if it's not what I'm expecting, then that immediately stands out as being weird and it draws my attention to it. \textbf{I would never have noticed if it were not plotted; I would never have thought to plot it.}''
\end{quoting}
Our participants described how these unexpected, serendipitous, discoveries were primarily facilitated by the auto-updating and automatic visualizations of \sys{} and made the system a valuable part of their workflow.

\section{Discussion and Future Work}
\label{sec: discussion}

Data science is messy.
There are a combinatorially large number of ways to slice a dataset, trying to find meaningful insights.
The goal of continuous data profiling is to augment a human's sense-making ability by automating the analysis feedback loop to be as fast as possible.
Previous work has established that automated systems can best facilitate data understanding by automating the need for manual specification \cite{Heer2019AgencyPA}.
We found that two different versions of automatic profiling help speed up this feedback loop in our user study.
Furthermore, we found evidence that the combination of automatic visualization, live updates, and code handoff leads to a smoother, more thorough analysis loop in our long-term deployment where our participants credited \sys{} with helping them find ``serendipitous discoveries'' in their dataset.

In real-world tasks, encouraging critical engagement is challenging because analysts must trade off finding insights and errors quickly with a thorough and exhaustive analysis of their data.
\sys{}'s design removes friction by saving time and clicks to better facilitate continuous data profiling.
Since \sys{} works with any pandas dataframe, users do not have to write or copy and paste profiling code that might be tightly coupled to a specific dataset.
This makes notebooks cleaner and easier to maintain.

Future tools can leverage the benefits of both code and automated visualization for data analysis through linked and deeply integrated data profiles.
Automatically presenting a starting set of profiling information and supporting follow-up analysis by enabling code exports helps reduce the feedback time during analysis.
This approach differs from other profiling systems that aim to include as much information as possible in the interface without handing off to code~\cite{Lee2021LuxAV, pandas-profiling}.

\subsection{Guiding users towards unknown insights}
Beyond making data analysis faster, automated systems like \sys{} can help users discover insights they might have otherwise missed.
These serendipitous discoveries present an interesting opportunity for tools to help users look at their data in new ways.
However, this process cannot be fully automated. 
Automatically presenting data profiles to users gives them the \textit{opportunity} to find insights.
Users must still take the time to look at and interpret if an insight or error is noteworthy.
Automated systems can augment human expertise, but do not replace it.
For example, in our user study, many participants missed important data quality issues like duplicate values, even though this information was readily available in either tool if one knew to check.
The most common types of unexpected errors discovered through \sys{} were strange distributions such as a totally flat distribution or weird frequent values.
The distribution information is very visually prominent in \sys{}, perhaps making it easier to discover in the interface.

Automated assistance in notebooks opens up the design space for further improvements toward guided analysis.
One exciting area for future work is the potential to integrate alerts into automatic data profiles to draw user attention to important errors.
For example, an alert could be displayed if a column has a number of null values or outliers greater than some threshold.
Alerts must be customizable and designed to minimize alert fatigue, or else a user may totally ignore them \cite{Shankar2022OperationalizingML}.
With existing inline, manual profilers~\cite{pandas-profiling} these alerts would be re-computed and displayed every time a user updates and re-profiles their data, quickly causing alert fatigue.
Tools like \sys{} present an opportunity for persistent alerts between profiles that can better support continuous data science.

\subsection{Authoring more analysis code for users}
Our export to code feature was very popular among participants, with many requests for even more ways to export to code.
Part of the benefit of \sys{}'s approach to exports is they are predictable: the system exports the same template code every time, with the dataframe and column names filled in. 
This is in contrast to generative approaches to code authoring such as Github Copilot~\cite{githubCopilot} where a model might produce different code for the same task depending on the prompt.
Users must then take time to understand this new code each time it is exported.
The downside to template approaches like ours is that it is less flexible for arbitrary analysis.

In our user study, we frequently observed participants needing to look up the documentation for how to write a certain command with the Pandas library, even if they were experienced users.
As tools continue to evolve to automatically write analysis code through text prompting, we think this will make data iteration even faster.
The linked, interactive outputs from systems like \sys{} becomes even more valuable to help users understand their data as the time it takes to write analysis code decreases, perhaps especially when users are not manually writing all of that code and need to understand its effect on their data.

\section{Conclusion}
In conclusion, we present \sys{}, a Jupyter notebook assistant that uses automatic, live, and linked data profiles to support continuous data profiling during data analysis.
In a controlled user study, we find users leverage two versions of our tool, dead or alive, to find the vast majority of insights during a data cleaning task.
Furthermore, we find that \sys{} easily fits into data scientists' real-world workflows and helps them discover unexpected insights in their data during a longitudinal case study.
We discuss how tools like \sys{} open up the design space for automated assistants to support continuous data profiling during analysis.
\acknowledgments{
We would like to thank Venkat Sivaraman, Katelyn Morrison, Alex Cabrera and the members of the Data Interaction Group at CMU for their feedback on this work; Hamilton Ulmer and the Rill Data team for the initial implementation of our data profiling charts; Wei Xu, Kevin Yager, and Esther Tsai at Brookhaven National Laboratory for their feedback and use of AutoProfiler.
This research was supported by Brookhaven National Laboratory through New York State funding and the Human-AI-Facility Integration (HAI-FI) initiative.
}
\bibliographystyle{abbrv-doi-hyperref}
\bibliography{refs}

\clearpage
\appendix 

\section*{Supplemental Materials}
\label{sec: supplemental materials}
Additional tables and figures about our user study and \inlinep{}.

\begin{table}[!h]
    \centering
    \begin{tabular}{rlllrl}
    \toprule
    PID & Condition & Analysis Freq & Pandas Freq & Years DS exp & Job \\
    \midrule
    P1  & AutoProfiler      & Monthly   & Monthly   & 3     & Data scientist  \\
    P2  & StaticProfiler    & Weekly    & Weekly    & 5     & Grad student  \\
    P3  & AutoProfiler      & Daily    & Daily      & 7     & Postdoc  \\
    P4  & StaticProfiler    & Weekly    & Monthly   & 4     & Grad student  \\
    P5  & AutoProfiler      & Daily    & Daily      & 2     & Data engineer  \\
    P6  & StaticProfiler    & Weekly    & Weekly    & 10     & Researcher  \\
    P7  & StaticProfiler    & Weekly    & Monthly   & 4     & Data journalist  \\
    P8  & StaticProfiler    & Daily    & Daily      & 3     & Data analyst  \\
    P9  & AutoProfiler      & Daily    & Daily      & 3     & Grad student  \\
    P10  & AutoProfiler     & Weekly    & Weekly    & 12     & Data journalist  \\
    P11  & StaticProfiler   & Weekly    & Weekly    & 4     & Data journalist  \\
    P12  & AutoProfiler     & Weekly    & Daily     & 5     & Data engineer  \\
    P13  & AutoProfiler     & Daily    & Daily      & 5     & Data engineer  \\
    P14  & StaticProfiler   & Daily    & Daily      & 3     & Data engineer  \\
    P15  & AutoProfiler     & Daily    & Weekly     & 2     & Grad student  \\
    P16  & StaticProfiler   & Weekly    & Weekly    & 5     & Grad student  \\
    \bottomrule
    \end{tabular}
    \caption{Background information on the participants for our user study.}
    \label{tab: participant details}
\end{table}
\begin{table}[!h]
    \centering
    \begin{tabular}{llrrr}
    \toprule
    \multicolumn{2}{c}{} & \multicolumn{3}{c}{\textit{Discovery Rate}} \\
    No. & Description & AP & SP & Avg. \\
    \midrule
    1 & Small number of missing values in county, beds, title                 & 50\% & 63\% & 56\% \\
    2 & Mostly missing in baths, sqft, description                            & 50\% & 63\% & 56\%\\
    3 & City has values that are lower and upper case                         & 75\% & 63\% & 69\%\\
    4 & Negative prices                                                       & 63\% & 75\% & 69\% \\
    5 & Date could be parsed to DateTime format                               & 63\% & 63\%  & 63\% \\
    6 & County has default values of \mintinline{Python}{"---"}               & 75\% & 88\%  & 81\% \\
    7 & Sqft has string values and should be converted to an int              & 63\% & 75\% & 69\%  \\
    8 & Outliers in sqft                                                      & 13\% & 0\% & 6\% \\
    9 & Outliers in price                                                     & 50\% & 38\% & 44\% \\
    10 & Duplicate datapoints (duplicate post\_ids)                           & 38\% & 38\% & 38\%  \\
    11 & Room\_in\_apt is almost all 0                                        & 75\% & 38\%  & 56\% \\
    12 & Dataset is only apartments in California                             & 13\% & 50\% & 31\% \\
    13 & Inspect correlations between any variable and price                  & 25\% & 0\% & 13\% \\
    \midrule
    14 & Data is not evenly distributed across years                          & 38\% & 38\%  & 38\% \\
    15 & Year and date column correspond (ensure consistency)                 & 38\% & 0\% & 19\%  \\
    16 & The price is not properly extracted from title for some rows         & 13\% & 0\% & 6\%  \\
    \bottomrule
    \end{tabular}
    \caption{Errors and insights repeated from \autoref{tab: errors in study} with further breakdown of discovery rate by condition. 
    The discovery rate of users with \sys{} (AP) and \inlinep{} (SP) along with the average across both conditions is reported. Since the same information is included in both tools, the discovery rates are largely comparable.
    }
    \label{tab: Split results}
\end{table}
\begin{figure*}[t]
    \centering
    \includegraphics[height=10cm, keepaspectratio]{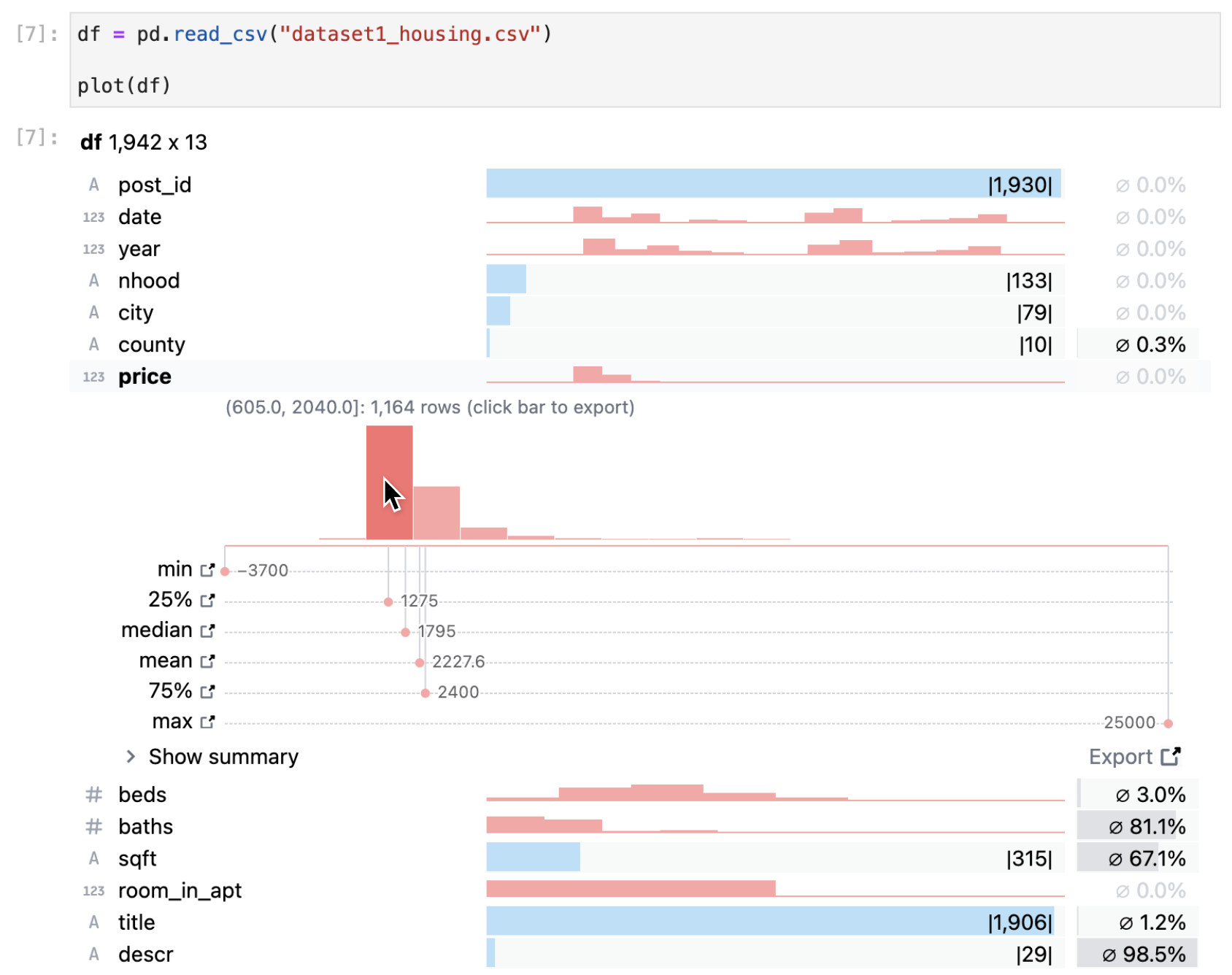}
    \caption{The \inlinep{} tool used in our user study. \inlinep{} shows the exact same information as \sys{}, however is presented inline and summoned by calling \mintinline{Python}{plot()} with a pandas dataframe. When the data updates, \mintinline{Python}{plot()} must be called again to re-visualize the new data whereas with \sys{} this update would happen automatically.}
    \label{fig:StaticProfiler}
\end{figure*}

\end{document}